\documentclass[aps,twocolumn,prl,showpacs,amsmath,amssymb,showkeys,10pt]{revtex4-1}

\usepackage{graphicx}
\usepackage{epstopdf}
\usepackage{braket}
\usepackage{hyperref}
\allowdisplaybreaks

\begin{document}

\title{Atom-field correlations in the weak-excitation limit of absorptive optical bistability}

\author{Th. K. Mavrogordatos}
\email[Email address(es): ]{themis.mavrogordatos@fysik.su.se; th.mavrogordatos@gmail.com}
\affiliation{Department of Physics, Stockholm University, SE-106 91, Stockholm, Sweden}

\date{\today}

\begin{abstract}
We calculate the steady-state and first-order time varying atom-field correlation functions in the weak-excitation limit of absorptive optical bistability from a linearized theory of quantum fluctuations. We formulate a Fokker-Planck equation in the positive $P$ representation following the phase-space analysis of [H. J. Carmichael, Phys. Rev. A {\bf 33}, 3262 (1986)] which is suitable for the determination of cross-correlations as it does not resort to adiabatic elimination. Special emphasis is placed on the limit of collective strong coupling as attained from a vanishing photon-loss rate. We compare to the cavity-transmission spectrum with reference to experimental results obtained for macroscopic dissipative systems, discussing the role of anomalous correlations arising as distinct nonclassical features.
\end{abstract}


\maketitle

Since the early development of quantum optics, a good number of quantum-statistical phenomena at the center of attention in the study of driven dissipative macroscopic systems have been associated with absorptive optical bistability which was theoretically predicted in 1969 \cite{Szoke1969}. For a low-Q cavity, or the {\it bad-cavity limit} \textemdash{calling} for the adiabatic elimination of the intracavity field \textemdash{a} linearized Fokker-Planck equation was derived in the Wigner representation and employed for the calculation of the second-order correlation function for the transmitted light, which demonstrated the existence of photon antibunching \cite{Casagrande1980, Walls1979}. Moreover, when the bistable absorber with dominant photon losses operates in the weak-excitation limit, the linewidth of the transmitted spectrum has been found to be much larger than the empty-cavity width and proportional to the atomic density, an evidence of atomic cooperativity \cite{Bonifacio1978}. Concurrently, the application of the spectral theorem for thermodynamic Green's functions in the low-temperature limit of the Dicke model produced the state equation of absorptive optical bistability, evidencing an explicit connection to superriadiance; the existence of the second-order superradiant phase transition in thermodynamic equilibrium and the existence of optical bistability as a first-order phase transition out of equilibrium were linked to the same basic matter-light interaction \cite{Bowden1979}.

A double-peaked Glauber-Sudarshan $P$ distribution in the steady state was demonstrated by \cite{Bonifacio1978}, and, alongside Agarwal and coworkers, entrenched the concept of a ``discontinuous formation of sidebands along the high-transmission branch'' of absorptive bistability \cite{Agarwal1978}. The calculation of the fluorescent spectrum in the bad-cavity limit of absorptive bistability showed a well-resolved Stark triplet along the upper branch without a discontinuous formation nor a linewidth narrowing of the central peak when distinguishing between like and unlike-atom correlations \cite{Carmichael1981}. This underlies the fundamental difference between the field scattered in the forwards direction and the output field generated as atomic fluorescence; for the latter there is no collective output channel. About a decade lapsed to see the demonstration of single-atom absorptive optical bistability in \cite{Savage1988} while the so-called {\it zero-system size} of absorptive bistability was linked to spontaneous symmetry breaking and dressed-state polarization in \cite{Alsing1991}. This critical effect, where quantum fluctuations cause switching between the dressed-state excitation ladders of the Jaynes-Cummings Hamiltonian, was experimentally demonstrated in \cite{Armen2009}. Corroborating its theoretical prediction, photon antibunching for the forwards-scattered light was observed for a small number of two-level atoms strongly coupled to a high-finesse cavity in \cite{Rempe1991}, while a violation of Schwarz inequality arising from the nonclassical correlations in the bunched light transmitted by an atomic ensemble was reported in \cite{Mielke1998}. Nonclassical features at weak excitation were also depicted in the intensity correlation functions of \cite{Foster2000}. 

Closer to our days, there has been a renewed interest in the behavior of atomic ensembles strongly coupled to modes supported by high-finesse cavities and exhibiting macroscopic coherence. The displacement of motionally coherent ultracold atoms trapped in a Fabry-Perot cavity provided access to both branches of dispersive optical bistability for very low intracavity excitation \cite{Gupta2007}. In the experiment of \cite{Brennecke2007}, the atomic ensemble oscillates between its ground state and a symmetric excited state in which a single excitation is shared by all the atoms, while the experiment of \cite{Gothe2019} employs a thermal cloud of Yb atoms in a high-Q cavity to focus on the upper branch of dispersive bistability with a high intracavity photon number. The size of cross-correlations for a few atoms strongly coupled to the intracavity field in absorptive optical bistability has been predicted to increase with the number of emitters in the upper branch \cite{Dombi2013}. Our very brief overview of many-atom bistability closes with the very recent proposal of \cite{Schachenmayer2020}, where a single emitter is surrounded by an atomic ensemble comprising coherently and dissipatively coupled emitters via dipolar interactions. The ensemble is then reduced to an effective active environment affecting the coupling of the single emitter to the cavity mode.

In this Letter, we apply a linear theory of quantum fluctuations in the phase-space representation in order to determine the light-matter correlation functions in the many-atom strong-coupling limit of absorptive optical bistability. When operating the bistable absorber in this particular regime, a Rabi doublet with narrowed peaks due to squeezing arises in the transmitted-light spectrum at weak excitation as a distinct signature of cooperative light-matter coupling \cite{Carmichael1986}. For our purpose, we need to formulate a procedure which does not rely on adiabatic elimination of system variables. After sketching a rudimentary road-map for transitioning from the master equation to the linearized Fokker-Planck equation for the fluctuations, we derive the equations of motion determining the evolution of the covariance matrix. We subsequently focus on a specific application of the linear theory to determine the first-order coherence properties of collective light-matter coupling before summarizing our results in view of accessing the analytically derived cross-correlations. 

Let us now start with the microscopic model underlying the macroscopic analysis of an optical resonator containing a homogeneously broadened two-level medium. We consider a system of $N$ homogeneously broadened two-level atoms, described by the operators of collective polarization, $J_{\pm} \equiv \sum_{j=1}^{N}\sigma_{j \pm}$, and inversion $J_z \equiv \sum_{j=1}^{N}\sigma_{jz}$ obeying the commutation relations of angular momentum [where $\sigma_{j \pm}, \sigma_{jz}$ are the pseudospin operators for the $j^{\rm th}$ atom], resonantly coupled to a coherently-driven cavity mode with frequency $\omega_0$; $a$ and $a^{\dagger}$ are the photon creation and annihilation operators, respectively. Each atom is radiatively damped by spontaneous emission to modes other than the privileged cavity mode with a damping rate $\gamma$, and the cavity field is damped by losses at partially reflecting mirrors with a photon dissipation rate $2\kappa$. Each atom is assumed to couple to the cavity mode with the same strength $g=[\omega_0 d^2/(2 \hbar \epsilon_0 V_Q)]^{1/2}$, where $d$ is the atomic dipole moment (independent of the atom's location), $\epsilon_0$ is the vacuum permittivity and $V_Q$ is the mode volume. In the familiar Born-Markov approximation, the system density operator $\rho$ obeys the Lindblad master equation (ME), known as the {\it master equation for optical bistability}  
\begin{align}\label{eq:MEbist}
\frac{d\rho}{dt}&=-i\frac{1}{2}\omega_0 [J_z, \rho]-i\omega_0[a^{\dagger}a, \rho] \notag \\
&+g[a^{\dagger}J_{-}-aJ_{+},\rho]-i[\bar{\mathcal{E}}_0 e^{-i\omega_0 t}a^{\dagger}+\bar{\mathcal{E}}_0^{*} e^{i\omega_0 t}a, \rho] \notag \\
&+\frac{\gamma}{2}\left(\sum_{j=1}^{N}2\sigma_{j-}\rho \sigma_{j+}-\frac{1}{2}J_{z}\rho - \frac{1}{2}\rho J_{z} -N\rho\right)\notag \\
&+\kappa(2a\rho a^{\dagger}-a^{\dagger}a\rho-\rho a^{\dagger}a),
\end{align}
where we have neglected thermal fluctuations \textemdash{this} is valid at optical frequencies even at room temperature, since $\hbar \omega_0/(k_B T) \gg 1$, but not for microwaves where the presence of thermal photons can be disregarded only at temperatures below a few mK. The interatomic distance is much larger than the resonant wavelength, $\lambda=2\pi c/\omega_0$, so that individual scattering records from different atoms may be obtained. The independent spontaneous emission channels, consistent with the Lindblad term in the third line of ME \eqref{eq:MEbist}, render the atoms {\it distinguishable}. 

Let us now discuss the equivalent phase-space representation to the evolution governed by the ME \eqref{eq:MEbist}. In the positive $P$ representation, the quasi-distribution function $P(\alpha, \alpha_{*},v,v_{*},m)$ is defined from the normally ordered characteristic function $\chi_N(\beta,\beta_{*},\xi,\xi_{*},\eta)$ \cite{Drummond1980, Carmichael1986PositiveP} via an integral relation involving the five {\it independent} complex variables $(\alpha, \alpha_{*}, v, v_{*},m)$ in a ten-dimensional space, as  
\begin{equation}\label{eq:Pdef}
\begin{aligned}
&\chi_N(\beta,\beta_{*},\xi,\xi_{*},\eta) \equiv {\rm tr}(\rho\, e^{i\beta_{*}a^{\dagger}} e^{i\beta a} e^{i\xi_{*}J_{+}}e^{i\eta J_z} e^{i\xi J_{-}})\\
&=\int d^2 \alpha \int d^2\alpha_{*} \int d^2v \int d^2v_{*} \int d^2 m \\
& \times P(\alpha, \alpha_{*},v,v_{*},m)e^{-i\beta_{*}\alpha_{*}}e^{-i\beta\alpha} e^{-i\xi_{*}v_{*}} e^{-i\xi v} e^{-i\eta m},
\end{aligned}
\end{equation}
where each integration extends over the entire complex plane. We define the following set of scaled phase-space variables as (see Sec. 15.2.1 of \cite{CarmichaelQO2})
\begin{subequations}\label{eq:scalingFP}
\begin{align}
i e^{-i\phi_0}\alpha&=n_{\rm s}^{1/2}\bar{\alpha}, \label{eq:scalingFP_a}\\
-i e^{i\phi_0}\alpha_{*}&=n_{\rm s}^{1/2}\bar{\alpha}_{*}, \label{eq:scalingFP_b}\\
i \sqrt{2}\, e^{-i\phi_0}v&=N \bar{v}, \label{eq:scalingFP_c}\\
-i \sqrt{2}\, e^{i\phi_0}v_{*}&=N \bar{v}_{*}, \label{eq:scalingFP_d}\\
m&=N\bar{m}, \label{eq:scalingFP_e}
\end{align}
\end{subequations}   
in which $n_{\rm s}\equiv \gamma^2/(8g^2)$ is the saturation photon number defining the weak-coupling limit of absorptive optical bistability ($n_{\rm s} \to \infty$ when $g/\gamma \to 0$), and the phase $\phi_0$ is introduced to ensure that the scaled driving-field amplitude $\bar{\mathcal{E}}_0$ (containing phase terms due to transmission from the cavity mirror and the dipole interaction) appears hereinafter as a real parameter.

In order to carry out the system-size expansion, we introduce fluctuations scaling with respect to the number of atoms such that the small-noise limit admitting a linearized treatment is attained for $N \gg 1$, the smallest parameter in the system. Based on \cite{Carmichael1986} and Sec. 15.2.1 of \cite{CarmichaelQO2} we write
\begin{subequations}\label{eq:scaleN}
\begin{align}
\bar{\alpha}&=\braket{\bar{a}(t)} + N^{-1/2}z, \label{eq:scaleN_a}\\
\bar{\alpha}_{*}&=\braket{\bar{a}^{\dagger}(t)} + N^{-1/2}z_{*}, \label{eq:scaleN_b}\\
\bar{v}&=\braket{\bar{J}_{-}(t)} + N^{-1/2} \nu, \label{eq:scaleN_c}\\
\bar{v}_{*}&=\braket{\bar{J}_{+}(t)} + N^{-1/2} \nu_{*}, \label{eq:scaleN_d}\\
\bar{m}&=\braket{\bar{J}_{z}(t)} + N^{-1/2}\mu, \label{eq:scaleN_e}
\end{align}
\end{subequations}
where $[\braket{\bar{a}(t)},\braket{\bar{a}^{\dagger}(t)}, \braket{\bar{J}_{-}(t)}, \braket{\bar{J}_{+}(t)}, \braket{\bar{J}_{z}(t)}]$ are the macroscopic averages of the operators defined in analogy to the scaling relations of Eqs. \eqref{eq:scalingFP}, satisfying the Maxwell-Bloch equations. 

The relation between the scaled intracavity amplitude $X \equiv \braket{\tilde{\bar{a}}}_{\rm ss}=\braket{\tilde{\bar{a}}^{\dagger}}_{\rm ss}$ (the tilde on top of operators and phase-space variables signifies a transformation to a frame rotating with the resonance frequency $\omega_0$, while the subscript ${\rm ss}$ denotes the steady state) and the scaled driving-field amplitude $Y \equiv |\bar{\mathcal{E}}_0|/(\kappa \sqrt{n_{\rm s}})$ is determined by the steady-state equation of absorptive optical bistability \cite{Carmichael1986}
\begin{equation}\label{eq:BistSE}
Y=X\left(1 + 2C \frac{1}{1+X^2}\right),
\end{equation}
where $2C \equiv 2N g^2/(\kappa\gamma)$ is the {\it cooperativity parameter}. For $C>4$, Eq. \eqref{eq:BistSE}  possesses two stable and one unstable solution. The two stable solutions define the {\it lower} and {\it upper} branch of bistability via $X<X_{-}$ and $X>X_{+}$, respectively, where $X_{\pm}\equiv (C-1) \pm \sqrt{C(C-4)}$ are the two turning points. In this work, we will be concerned only with the lower branch, and specifically with very low excitation amplitudes such that the input-output curve of absorptive optical bistability can be approximated by a straight line $Y \approx (1+2C) X$ with a slope deviating substantially from unity since we are dealing with a cooperativity parameter $2C > 8$. From the Maxwell-Bloch equations in the steady state, one obtains for the collective atomic polarization and inversion,
\begin{equation}\label{eq:atomicPZ}
\braket{\tilde{\bar{J}}_{-}}_{\rm ss}=\braket{\tilde{\bar{J}}_{+}}_{\rm ss}=-\frac{X}{1 + X^2}, \quad \braket{\bar{J}_{z}}_{\rm ss}=-\frac{1}{1+X^2}.
\end{equation}
In the weak-excitation limit, the total spontaneous emission rate is \cite{CarmichaelQO2}
\begin{equation}
 R_{\gamma}\equiv \frac{1}{2}\gamma(N+\braket{J_z}_{\rm ss})=\frac{1}{2}\gamma N (1+\braket{\bar{J}_{z}}_{\rm ss}) \approx \frac{1}{2}\gamma N X^2,
\end{equation}
while the rate of cavity emissions is 
\begin{equation}
R_{\kappa}\equiv 2\kappa \braket{a^{\dagger}a}_{\rm ss}= 2\kappa n_{\rm s}\braket{\tilde{\bar{a}}^{\dagger}\tilde{\bar{a}}}_{\rm ss} \approx  2\kappa n_{\rm s} X^2,
\end{equation}
where we have used the operator scaling corresponding to Eqs. \eqref{eq:scalingFP} and omitted corrections of order $X^4$. The ratio of these two rates is $R_{\gamma}/R_{\kappa}=2C$, whence the existence of bistability sets spontaneous emission as the dominant decoherence channel. 

The mean-field equations arise as terms set to zero by giving contributions of order $N^{1/2}$ in the full Fokker-Planck equation (FPE) for $P(\alpha, \alpha_{*},v,v_{*},m)$ corresponding to the ME \eqref{eq:MEbist} and subject to the system-size expansion of Eqs. \eqref{eq:scaleN}. Subsequently, we drop terms of order $N^{-1/2}$ to derive an equation for the distribution of fluctuations about the macroscopic motion (see the procedure detailed in Sec. 15.2.1 of \cite{CarmichaelQO2}). The scaled quasi-distribution function, 
\begin{equation}
\begin{aligned}
&\tilde{\bar{P}}(\tilde{z}, \tilde{z}_{*}, \tilde{\nu}, \tilde{\nu}_{*},\mu)\equiv \frac{1}{2}n_{\rm s}N^{1/2}\\
& \times P(\tilde{\bar{\alpha}}(\tilde{z},t), \tilde{\bar{\alpha}}_{*}(\tilde{z}_{*},t), \tilde{\bar{v}}(\tilde{\nu},t), \tilde{\bar{v}}_{*}(\tilde{\nu}_{*},t), m(\mu,t)),
\end{aligned}
\end{equation}
satisfies the linearized FPE
\begin{equation}\label{eq:FPE}
\frac{\partial \tilde{\bar{P}}}{\partial t}=\left(-\tilde{\boldsymbol{Z}}^{\prime \top} \bar{\boldsymbol{J}}_{\rm ss} \tilde{\boldsymbol{Z}} + \frac{1}{2}\tilde{\boldsymbol{Z}}^{\prime \top} \bar{\boldsymbol{D}}_{\rm ss} \tilde{\boldsymbol{Z}}^{\prime} \right) \tilde{\bar{P}},
\end{equation}
with
\begin{equation}\label{eq:defvectors}
\tilde{\boldsymbol{Z}} \equiv \begin{pmatrix}
\tilde{z} \\ \tilde{z}_{*} \\ \tilde{\nu} \\ \tilde{\nu}_{*} \\ \tilde{\mu}
\end{pmatrix}, \quad \quad \tilde{\boldsymbol{Z}}^{\prime} \equiv \begin{pmatrix}
\partial / \partial\tilde{z} \\ \partial / \partial\tilde{z}_{*} \\ \partial / \partial\tilde{\nu} \\ \partial / \partial\tilde{\nu}_{*} \\ \partial / \partial\tilde{\mu}
\end{pmatrix}.
\end{equation}
In the linearized FPE \eqref{eq:FPE}, $\bar{\boldsymbol{J}}_{\rm ss}$ is the Jacobian matrix and $\bar{\boldsymbol{D}}_{\rm ss}$ is the diffusion matrix, both having real constants as their matrix elements (i.e., not depending on the phase-space variables). In the positive $P$ representation, these two matrices assume the form
\begin{equation}\label{eq:JacobFull}
\bar{\boldsymbol{J}}_{\rm ss}=\frac{\gamma}{2}\begin{pmatrix}
-\xi & 0 & \xi 2C & 0 & 0 \\
0 & -\xi & 0 & \xi 2C & 0 \\
-\frac{1}{1+X^2} & 0 & -1 & 0 & X \\
0 & -\frac{1}{1+X^2} & 0 & -1 & X \\
\frac{X}{1+X^2} & \frac{X}{1+X^2} & -X & -X & -2
\end{pmatrix}
\end{equation}
and
\begin{equation}\label{eq:DiffM}
\bar{\boldsymbol{D}}_{\rm ss}=\gamma\frac{X^2}{1+X^2} \begin{pmatrix}
0 & 0 & 0 & 0 & 0\\
0 & 0 & 0 & 0 & 0\\
0 & 0 & -1 & 0 & 0\\
0 & 0 & 0 & 1 & 0\\
0 & 0 & 0 & 0 & 4
\end{pmatrix},
\end{equation}
respectively, with $\xi \equiv 2\kappa/\gamma$ the dimensionless ratio of the two decay rates whose relation to $C$ defines distinct regions of operation for the bistable absorber. The diffusion matrix of Eq. \eqref{eq:DiffM} is manifestly nonpositive semidefinite. In the linearized theory of fluctuations under current consideration, however, steady-state moments and the spectrum of fluctuations can be calculated in the original five-dimensional space through a naive application of familiar formal expressions to a Fokker-Planck equation in the Glauber-Sudarshan $P$-representation with a nonpositive-definite diffusion. In other words, we carry on with the calculations overlooking the fact that the diffusion matrix is not positive semidefinite (an assumption which is not permissible in principle for a nonlinear FPE).

The covariance matrix corresponding to the vector $\tilde{\boldsymbol{Z}}(t)$ is defined with respect to $\tilde{\bar{P}}(\tilde{z}, \tilde{z}_{*}, \tilde{\nu}, \tilde{\nu}_{*},\mu)$ as
\begin{equation}\label{eq:defCovMatrix}
\boldsymbol{C}_{\rm ss}(\tau) \equiv \lim_{t \to \infty}\left(\overline{\tilde{\boldsymbol{Z}}(t)\tilde{\boldsymbol{Z}}^{\top}(t+\tau)}\right)_{\tilde{\bar{P}}}.
\end{equation}
It obeys the equation of motion
\begin{equation}\label{eq:eomC}
\frac{d\boldsymbol{C}_{\rm ss}}{d\tau}=\begin{cases}\boldsymbol{C}_{\rm ss} \bar{\boldsymbol{J}}_{\rm ss}^{\top} \quad \quad \tau > 0 \\
\bar{\boldsymbol{J}}_{\rm ss}\boldsymbol{C}_{\rm ss} \quad  \quad \tau < 0 \end{cases},
\end{equation}
while at $\tau=0$,
\begin{equation}\label{eq:initialcondC}
\bar{\boldsymbol{J}}_{\rm ss} \boldsymbol{C}_{\infty} + \boldsymbol{C}_{\infty} \bar{\boldsymbol{J}}_{\rm ss} = -\bar{\boldsymbol{D}}_{\rm ss}.
\end{equation} 
We extract the atom-field correlation functions from the vector
\begin{equation}\label{eq:vectorCovM}
\begin{pmatrix}
C_{\rm ss}^{\tilde{\nu}_{*}\tilde{z}}(\tau) \\
C_{\rm ss}^{\tilde{\nu}_{*}\tilde{z}_{*}}(\tau)\\
C_{\rm ss}^{\tilde{\nu}_{*}\tilde{\nu}}(\tau)\\
C_{\rm ss}^{\tilde{\nu}_{*}\tilde{\nu}_{*}}(\tau)\\
C_{\rm ss}^{\tilde{\nu}_{*}\mu}(\tau)
\end{pmatrix}
=N \lim_{t\to\infty} \begin{pmatrix}
\braket{\Delta \tilde{\bar{J}}_{+}(t) \Delta\tilde{\bar{a}}(t+\tau)}\\
\braket{\Delta \tilde{\bar{J}}_{+}(t) \Delta\tilde{\bar{a}}^{\dagger}(t+\tau)}\\
\braket{\Delta \tilde{\bar{J}}_{+}(t) \Delta\tilde{\bar{J}}_{-}(t+\tau)}\\
\braket{\Delta \tilde{\bar{J}}_{+}(t) \Delta\tilde{\bar{J}}_{+}(t+\tau)}\\
\braket{\Delta \tilde{\bar{J}}_{+}(t) \Delta\bar{J}_{z}(t+\tau)}
\end{pmatrix},
\end{equation}
which is the fourth row of the covariance matrix. From Eq. \eqref{eq:eomC}, we find that this vector obeys the equation of motion
\begin{equation}
\frac{d}{d\tau}\begin{pmatrix}
C_{\rm ss}^{\tilde{\nu}_{*}\tilde{z}} \\
C_{\rm ss}^{\tilde{\nu}_{*}\tilde{z}_{*}}\\
C_{\rm ss}^{\tilde{\nu}_{*}\tilde{\nu}}\\
C_{\rm ss}^{\tilde{\nu}_{*}\tilde{\nu}_{*}}\\
C_{\rm ss}^{\tilde{\nu}_{*}\mu}
\end{pmatrix} = \bar{\boldsymbol{J}}_{\rm ss} \begin{pmatrix}
C_{\rm ss}^{\tilde{\nu}_{*}\tilde{z}}\\
C_{\rm ss}^{\tilde{\nu}_{*}\tilde{z}_{*}}\\
C_{\rm ss}^{\tilde{\nu}_{*}\tilde{\nu}}\\
C_{\rm ss}^{\tilde{\nu}_{*}\tilde{\nu}_{*}}\\
C_{\rm ss}^{\tilde{\nu}_{*}\mu}
\end{pmatrix}.
\end{equation}
The corresponding initial conditions are contained in the steady-state covariance matrix $\boldsymbol{C}_{\infty} \equiv\boldsymbol{C}_{\rm ss}(0)$.

Approximating the Jacobian matrix of Eq. \eqref{eq:JacobFull} in the weak-excitation limit, $X \ll X_{-}$, the equations of motion for the various correlation-vector components read
\begin{subequations}\label{eq:system}
\begin{align}
\frac{d C_{\rm ss}^{\tilde{\nu}_{*}\tilde{z}}}{d\bar{\tau}}&=-\xi C_{\rm ss}^{\tilde{\nu}_{*}\tilde{z}} + \xi 2 C\, C_{\rm ss}^{\tilde{\nu}_{*}\tilde{\nu}}, \label{eq:system_a}\\
\frac{d C_{\rm ss}^{\tilde{\nu}_{*}\tilde{z}_{*}}}{d\bar{\tau}}&=-\xi C_{\rm ss}^{\tilde{\nu}_{*}\tilde{z}_{*}} + \xi 2 C\, C_{\rm ss}^{\tilde{\nu}_{*}\tilde{\nu}_{*}}, \label{eq:system_b}\\
\frac{d C_{\rm ss}^{\tilde{\nu}_{*}\tilde{\nu}}}{d\bar{\tau}}&=-C_{\rm ss}^{\tilde{\nu}_{*}\tilde{\nu}}-C_{\rm ss}^{\tilde{\nu}_{*}\tilde{z}} + X C_{\rm ss}^{\tilde{\nu}_{*}\mu}, \label{eq:system_c}\\
\frac{d C_{\rm ss}^{\tilde{\nu}_{*}\tilde{\nu}_{*}}}{d\bar{\tau}}&=-C_{\rm ss}^{\tilde{\nu}_{*}\tilde{\nu}_{*}}-C_{\rm ss}^{\tilde{\nu}_{*}\tilde{z}_{*}} + X C_{\rm ss}^{\tilde{\nu}_{*}\mu}, \label{eq:system_d}\\
\frac{d C_{\rm ss}^{\tilde{\nu}_{*}\mu}}{d\bar{\tau}}&=-2C_{\rm ss}^{\tilde{\nu}_{*}\mu} + X (C_{\rm ss}^{\tilde{\nu}_{*}\tilde{z}}+C_{\rm ss}^{\tilde{\nu}_{*}\tilde{z}_{*}})\notag\\
&-X(C_{\rm ss}^{\tilde{\nu}_{*}\tilde{\nu}}+C_{\rm ss}^{\tilde{\nu}_{*}\tilde{\nu}_{*}}), \label{eq:system_e}
\end{align}
\end{subequations}
where $\bar{\tau}\equiv \gamma \tau/2$ is the dimensionless time. The initial conditions in the weak-excitation limit follow from Eq. \eqref{eq:initialcondC} as \cite{Carmichael1986} 
\begin{subequations}\label{eq:initialcond}
\begin{align}
&C_{\rm ss}^{\tilde{\nu}_{*}\tilde{z}}(0)=X^4\frac{\xi 2C(2+\xi+2C)}{(1+2C)^2 (\xi+1)^2}, \label{eq:initialcond_a} \\
&C_{\rm ss}^{\tilde{\nu}_{*}\tilde{z}_{*}}(0)=-X^2 \frac{\xi 2C}{(1+2C)(\xi+1)}, \label{eq:initialcond_b} \\
&C_{\rm ss}^{\tilde{\nu}_{*}\tilde{\nu}}(0)=X^4 \frac{2C(2+\xi+2C)+(\xi+1)^2}{(1+2C)^2(\xi+1)^2}, \label{eq:initialcond_c} \\
&C_{\rm ss}^{\tilde{\nu}_{*}\tilde{\nu}_{*}}(0)=-X^2\frac{1+2C+2\xi}{(\xi+1)(1+2C)}, \label{eq:initialcond_d} \\
&C_{\rm ss}^{\tilde{\nu}_{*}\mu}(0)=X^3 \frac{2C+\xi+1}{(1+2C)(\xi+1)}.\label{eq:initialcond_e}
\end{align}
\end{subequations}
\begin{figure}[!ht]
\begin{center}
\includegraphics[width=0.5\textwidth]{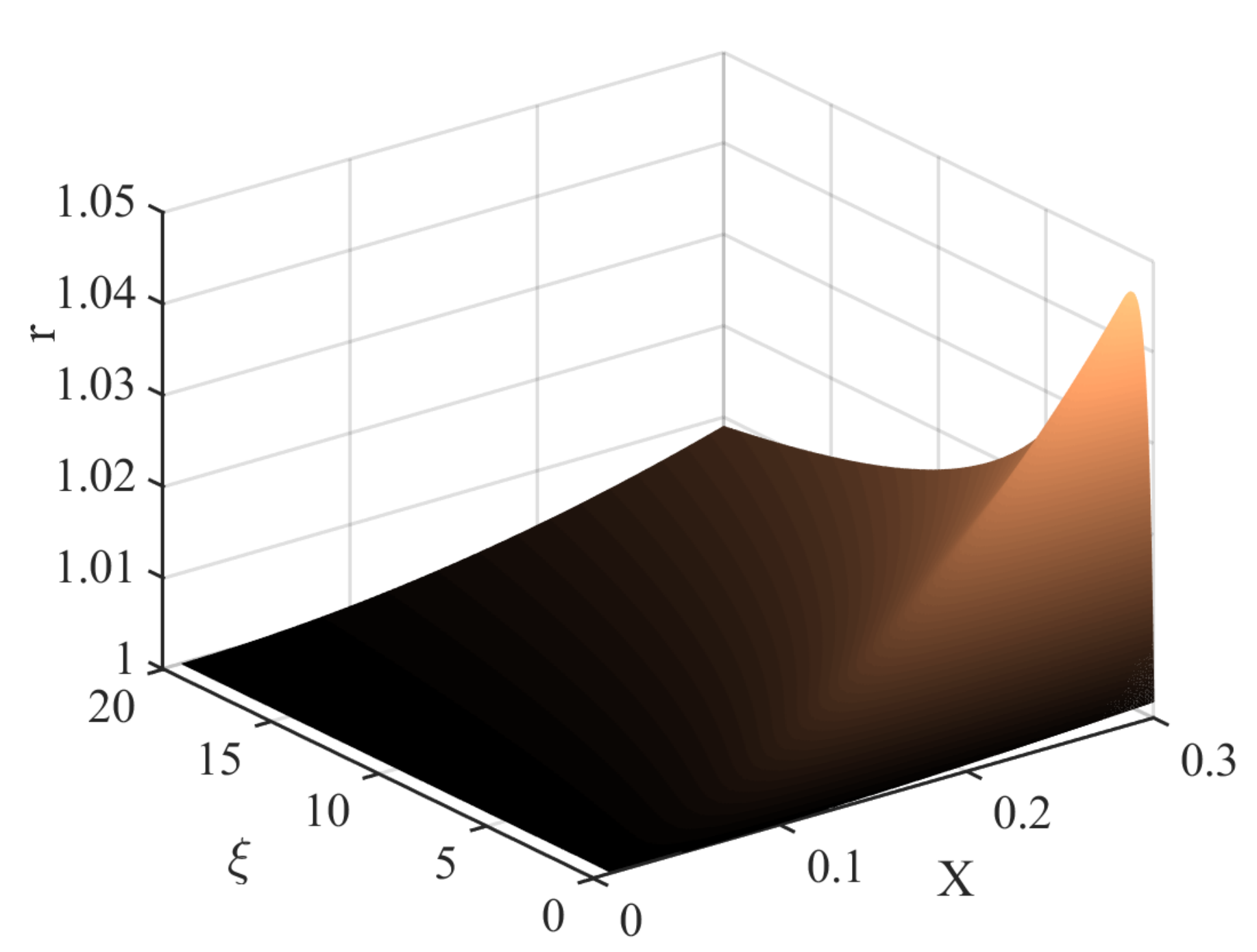}
\end{center}
\caption{{\it Ratio of steady-state averages with corrections of order $X^4$.} The ratio $r=r(X,\xi)=|\braket{\tilde{\bar{a}}^{\dagger}\tilde{\bar{a}}}_{\rm ss}/\braket{\tilde{\bar{J}}_{+}\tilde{\bar{a}}}_{\rm ss}|$ is depicted as a function of the intracavity amplitude $X$ and the parameter $\xi$ for $N=C=100$.}
\label{fig:ratio}
\end{figure}
From Eqs. \eqref{eq:initialcond_a} and \eqref{eq:initialcond_b}, we can calculate the steady-state averages
\begin{equation}\label{eq:CorrSS1}
\begin{aligned}
&\braket{ \tilde{\bar{J}}_{+}\tilde{\bar{a}}}_{\rm ss}=\braket{\tilde{\bar{J}}_{+}}_{\rm ss}\braket{\tilde{\bar{a}}}_{\rm ss}+\frac{1}{N}C_{\rm ss}^{\tilde{\nu}_{*}\tilde{z}}(0)\\
&\approx -X^2 \left[1-X^2\frac{\xi 2C(2+\xi+2C)}{N(1+2C)^2 (\xi+1)^2}\right]
\end{aligned}
\end{equation}
and
\begin{equation}\label{eq:CorrSS2}
\begin{aligned}
&\braket{ \tilde{\bar{J}}_{+}\tilde{\bar{a}}^{\dagger}}_{\rm ss}=\braket{\tilde{\bar{J}}_{+}}_{\rm ss}\braket{\tilde{\bar{a}}^{\dagger}}_{\rm ss}+\frac{1}{N}C_{\rm ss}^{\tilde{\nu}_{*}\tilde{z}_{*}}(0)\\
&\approx -X^2 \left[1+\frac{\xi 2C}{N(1+2C)(\xi+1)}\right],
\end{aligned}
\end{equation}
demonstrating explicit corrections of order $N^{-1} \ll 1$ arising from the linearized treatment of quantum fluctuations. The ratio of the latter to the former steady-state average tends to zero in the good-cavity limit ($\xi \ll 1$) and asymptotically approaches $(N+1)/N$ (with $N \gg 1$, as required for the small-noise analysis to hold) in the bad-cavity limit ($\xi \gg 2C$), exhibiting a negligible variation across the entire range of $\xi$. We note that the correction in the average of Eq. \eqref{eq:CorrSS1} is further reduced due to the presence of the $X^2$ term. Since the steady-state covariance matrix $\boldsymbol{C}_{\infty}$ is symmetric and has real elements for purely absorptive bistability [see also Eqs. (15.120c)-(15.120d) of \cite{CarmichaelQO2}], we have
\begin{equation}
\begin{aligned}
\braket{\Delta \tilde{\bar{J}}_{+}\Delta\tilde{\bar{a}}}_{\rm ss}=\braket{\Delta\tilde{\bar{a}}\Delta \tilde{\bar{J}}_{+}}_{\rm ss}&=\braket{\Delta\tilde{\bar{a}}^{\dagger}\Delta \tilde{\bar{J}}_{-}}_{\rm ss}\\
&=\braket{\Delta \tilde{\bar{J}}_{-}\Delta\tilde{\bar{a}}^{\dagger}}_{\rm ss}
\end{aligned}
\end{equation}
and
\begin{equation}
\begin{aligned}
\braket{\Delta \tilde{\bar{J}}_{+}\Delta\tilde{\bar{a}}^{\dagger}}_{\rm ss}=\braket{\Delta\tilde{\bar{a}}^{\dagger}\Delta \tilde{\bar{J}}_{+}}_{\rm ss}&=\braket{\Delta\tilde{\bar{a}}\Delta \tilde{\bar{J}}_{-}}_{\rm ss}\\
&=\braket{\Delta \tilde{\bar{J}}_{-}\Delta\tilde{\bar{a}}}_{\rm ss}.
\end{aligned}
\end{equation}

From Eq. (15.103b) of \cite{CarmichaelQO2} we read that
\begin{equation}
 \braket{\Delta \tilde{\bar{a}}^{\dagger}\Delta\tilde{\bar{a}}}_{\rm ss}\approx N^{-1} X^4\, 2C  \frac{\xi 2C (2+\xi+2C)}{(1+2C)^2(\xi+1)^2},
\end{equation}
yielding
\begin{equation}\label{eq:photonav}
  \braket{\tilde{\bar{a}}^{\dagger}\tilde{\bar{a}}}_{\rm ss}=X^2 \left[1 + X^2  2C  \frac{\xi 2C (2+\xi+2C)}{N(1+2C)^2(\xi+1)^2} \right],
\end{equation}
in the weak-excitation limit. Hence,
\begin{equation}
 \frac{\braket{\Delta \tilde{\bar{a}}^{\dagger}\Delta\tilde{\bar{a}}}_{\rm ss}}{\braket{\Delta \tilde{\bar{J}}_{+}\Delta\tilde{\bar{a}}}_{\rm ss}}=2C,
\end{equation}
which reveals the role of atomic cooperativity along the lower branch of absorptive bistability. In Fig. \ref{fig:ratio}, we plot the ratio of the scaled steady-state values given by Eqs. \eqref{eq:photonav} and \eqref{eq:CorrSS1}, which is an explicit function of the intracavity amplitude $X$. The largest deviation of the quantity $r(X, \xi)\equiv |\braket{\tilde{\bar{a}}^{\dagger}\tilde{\bar{a}}}_{\rm ss}/\braket{\tilde{\bar{J}}_{+}\tilde{\bar{a}}}_{\rm ss}|$ from unity occurs for $\xi=1$ as a consequence of impedance matching for the two decoherence channels (see also Sec. V of \cite{Carmichael1986}). 

We now proceed with the solution of the equations of motion \eqref{eq:system} to produce the desired first-order correlation functions. We retain terms of the same order in $X$ on the right-hand side of the equations comprising the system \eqref{eq:system}, in order to match the left-hand side as read from the initial conditions \eqref{eq:initialcond}. This means that we drop the term $X C_{\rm ss}^{\tilde{\nu}_{*}\mu}$ from Eq. \eqref{eq:system_c} together with the terms $C_{\rm ss}^{\tilde{\nu}_{*}\tilde{z}}$, $C_{\rm ss}^{\tilde{\nu}_{*}\tilde{\nu}}$ from Eq. \eqref{eq:system_e}. The transformed equations are then consistently reorganized in the following three subsets:
\begin{subequations}\label{eq:LT}
\begin{align}
&\begin{pmatrix}
\xi + \bar{s} & -\xi 2C \\
1 & 1+\bar{s}
\end{pmatrix} \begin{pmatrix}
\bar{\mathcal{C}}_{\rm ss}^{\tilde{\nu}_{*}\tilde{z}}(\bar{s}) \\
\bar{\mathcal{C}}_{\rm ss}^{\tilde{\nu}_{*}\tilde{\nu}}(\bar{s})
\end{pmatrix}=\notag \\
& \begin{pmatrix}
C_{\rm ss}^{\tilde{\nu}_{*}\tilde{z}}(0) \\ C_{\rm ss}^{\tilde{\nu}_{*}\tilde{\nu}}(0)
\end{pmatrix}+ X \bar{\mathcal{C}}_{\rm ss}^{\tilde{\nu}_{*}\mu} \begin{pmatrix}
0 \\ 1
\end{pmatrix}, \label{eq:LT_a} \\
&\begin{pmatrix}
\xi + \bar{s} & -\xi 2C \\
1 & 1+\bar{s}
\end{pmatrix} \begin{pmatrix}
\bar{\mathcal{C}}_{\rm ss}^{\tilde{\nu}_{*}\tilde{z}_{*}}(\bar{s}) \\
\bar{\mathcal{C}}_{\rm ss}^{\tilde{\nu}_{*}\tilde{\nu}_{*}}(\bar{s})
\end{pmatrix}=\begin{pmatrix}
C_{\rm ss}^{\tilde{\nu}_{*}\tilde{z}_{*}}(0) \\ C_{\rm ss}^{\tilde{\nu}_{*}\tilde{\nu}_{*}}(0)
\end{pmatrix}, \label{eq:LT_b} \\
&(2+\bar{s}) \bar{\mathcal{C}}_{\rm ss}^{\tilde{\nu}_{*}\mu}(\bar{s})=C_{\rm ss}^{\tilde{\nu}_{*}\mu}(0) +X[\bar{\mathcal{C}}_{\rm ss}^{\tilde{\nu}_{*}\tilde{z}_{*}}(\bar{s})-\bar{\mathcal{C}}_{\rm ss}^{\tilde{\nu}_{*}\tilde{\nu}_{*}}(\bar{s})], \label{eq:LT_c}
\end{align}
\end{subequations}
in which we have introduced the scaled quantities:
\begin{equation}\label{eq:defLTv}
\bar{s}\equiv 2s/\gamma, \quad \mathcal{C}_{\rm ss}^{ij}(s)=\frac{2}{\gamma}\bar{\mathcal{C}}_{\rm ss}^{ij}(\bar{s}),
\end{equation}
where $\mathcal{C}_{\rm ss}^{ij}(\bar{s})$ is the Laplace transform of $C^{ij}_{\rm ss}(\tau)$, $i,j=\tilde{z}, \tilde{z}_{*}, \tilde{\nu}, \tilde{\nu}_{*}, \mu$. 
\begin{figure}
\begin{center}
\includegraphics[width=0.45\textwidth]{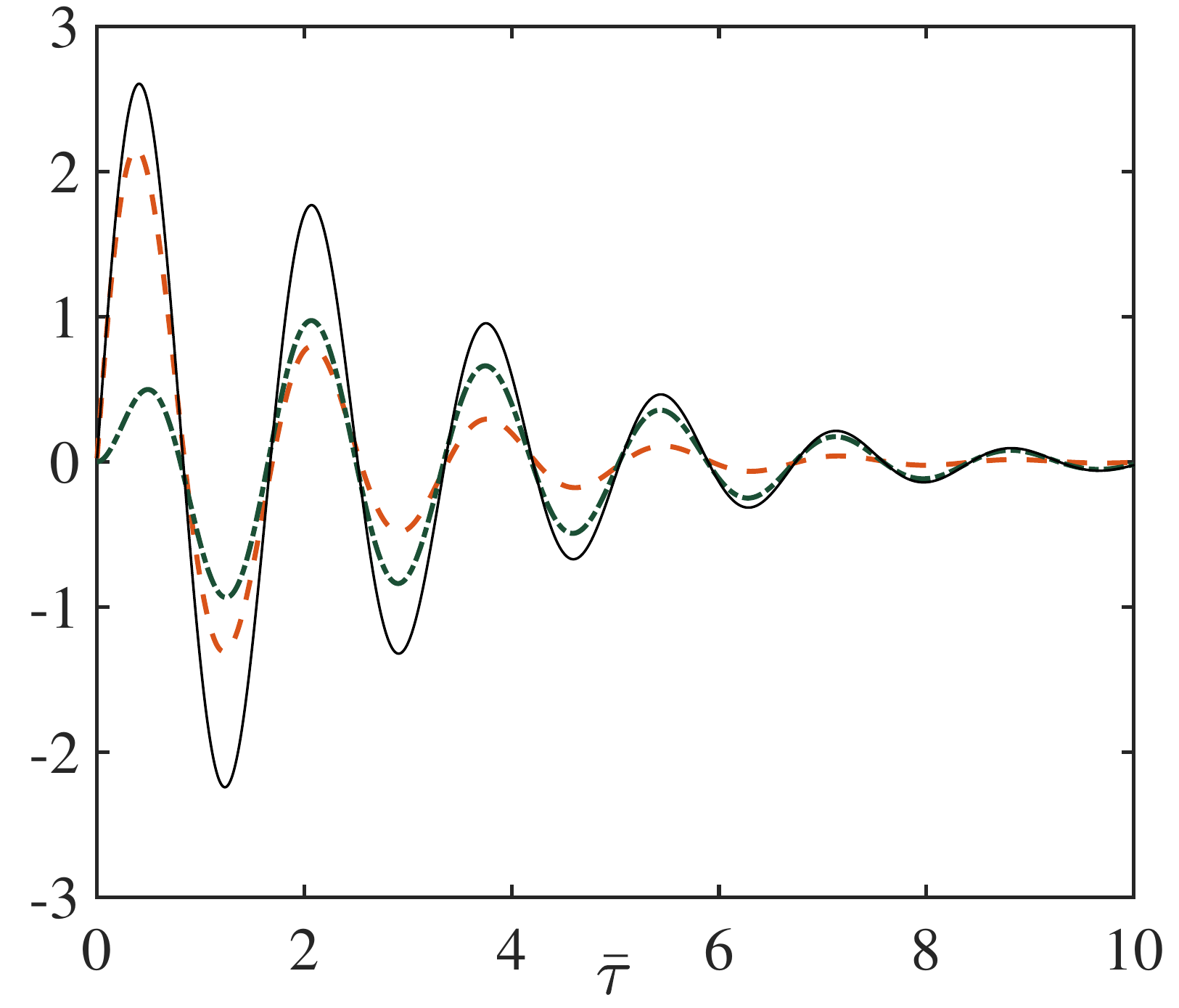}
\end{center}
\caption{{\it The two components of $C_{\rm ss}^{\tilde{\nu}_{*}\tilde{z}}(\bar{\tau})$ in the collective strong-coupling regime.} The component $C_{\rm ss}^{\tilde{\nu}_{*}\tilde{z};1}(\bar{\tau})/X^4$ from Eq. \eqref{eq:1stcondR} is plotted against the dimensionless time-delay $\bar{\tau}=\gamma \tau /2$ in a dashed orange line, the component $C_{\rm ss}^{\tilde{\nu}_{*}\tilde{z};2}(\bar{\tau})/X^4$ from Eq. \eqref{eq:2ndcondR} is plotted in a dot-dash green line, while their sum, $C_{\rm ss}^{\tilde{\nu}_{*}\tilde{z}}(\bar{\tau})$, is plotted in solid black for $(g,\kappa, \gamma)/2\pi=(1.06, 0.88, 10)\,$MHz, $N=310$ (parameters used in the experiment by Raizen and coworkers \cite{Raizen1989} to give $C \approx 40$ and $\xi \approx 0.18$, observing the typical hierarchy of scales $g\sqrt{N}>\gamma>2\kappa$).}
\label{fig:components}
\end{figure}

From the subset of Eqs. \eqref{eq:LT_a}, following the inversion of the matrix on the left-hand side, we can write
\begin{equation}\label{eq:systemCorr}
\begin{aligned}
& \begin{pmatrix}
\bar{\mathcal{C}}_{\rm ss}^{\tilde{\nu}_{*}\tilde{z}}(\bar{s}) \\
\bar{\mathcal{C}}_{\rm ss}^{\tilde{\nu}_{*}\tilde{\nu}}(\bar{s})
\end{pmatrix}=\frac{1}{(\xi +\bar{s})(1+\bar{s})+\xi 2C}\\
& \times\begin{pmatrix}
1 + \bar{s} & \xi 2C \\
-1 & \xi+\bar{s}
\end{pmatrix} \left[\begin{pmatrix}
C_{\rm ss}^{\tilde{\nu}_{*}\tilde{z}}(0) \\ C_{\rm ss}^{\tilde{\nu}_{*}\tilde{\nu}}(0)
\end{pmatrix}+ X \bar{\mathcal{C}}_{\rm ss}^{\tilde{\nu}_{*}\mu} \begin{pmatrix}
0 \\ 1
\end{pmatrix}\right],
\end{aligned}
\end{equation}
whence we find for the first correlation function of interest,
\begin{equation}\label{eq:Laplace1f}
\begin{aligned}
\bar{\mathcal{C}}_{\rm ss}^{\tilde{\nu}_{*}\tilde{z}}(\bar{s})&=\frac{(1+\bar{s})C_{\rm ss}^{\tilde{\nu}_{*}\tilde{z}}(0) + \xi 2CC_{\rm ss}^{\tilde{\nu}_{*}\tilde{\nu}}(0)}{(\xi +\bar{s})(1+\bar{s})+\xi 2C} \\
&+ \frac{\xi 2C X \bar{\mathcal{C}}_{\rm ss}^{\tilde{\nu}_{*}\mu}(\bar{s})}{(\xi +\bar{s})(1+\bar{s})+\xi 2C},
\end{aligned}
\end{equation}
where 
\begin{equation}
\bar{\mathcal{C}}_{\rm ss}^{\tilde{\nu}_{*}\mu}(\bar{s})=\frac{X^3}{(1+2C)(\xi+1)} \frac{(\xi+\bar{s})(\xi+1)+2C\bar{s}}{(\xi+\bar{s})(1+\bar{s})+\xi 2C}.
\end{equation}
\begin{figure}
\centering
\includegraphics[width=0.45\textwidth]{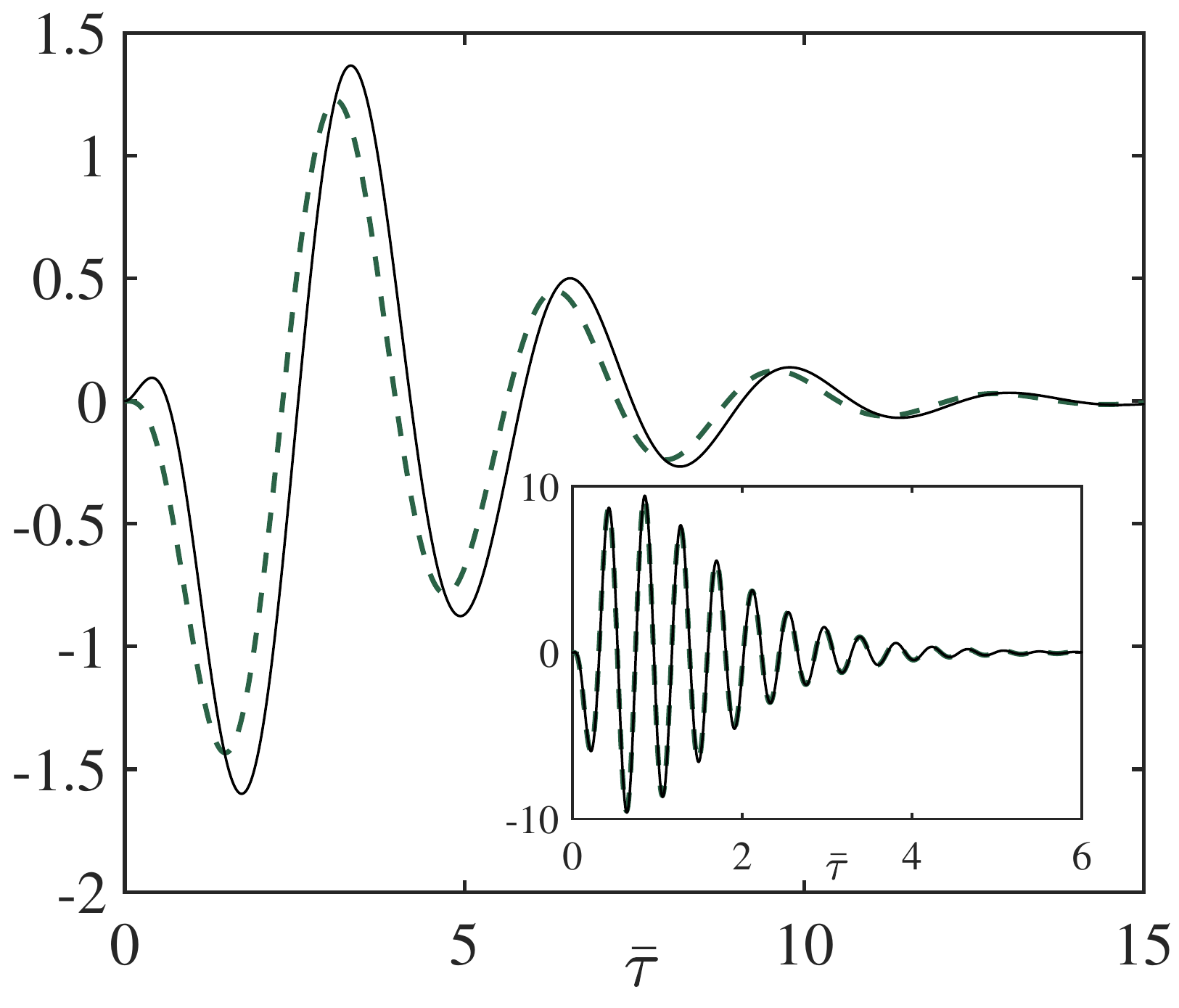}
\caption{{\bf Fig. 3} {\it Correlations with a squared Lorentzian distribution.} The sum of the two components: $C_{\rm ss}^{\tilde{\nu}_{*}\tilde{z};2}(\bar{\tau})/X^4$ obtained from Eq. \eqref{eq:2ndcondR}, and $C_{\rm ss}^{\tilde{z}_{*}\tilde{z};2}(\bar{\tau})/X^4$ obtained from Eq. \eqref{eq:trlight2R}, is plotted in a solid black line, superimposed on the component $C_{\rm ss}^{\tilde{z}_{*}\tilde{z};2}(\bar{\tau})/X^4$ alone, depicted by a dashed green line. Parameters: $\xi=0.05$, $C=40$. The inset depicts the same quantities, but for $\xi \approx 1.9$, $C \approx 58$, corresponding to the parameters used for the experiment of Mielke and collaborators (see Fig. 4 of \cite{Mielke1998}).}
\label{fig:compRes}
\end{figure}
We then separate the Laplace transform in Eq. \eqref{eq:Laplace1f} as
\begin{equation}\label{eq:separation}
\bar{\mathcal{C}}_{\rm ss}^{\tilde{\nu}_{*}\tilde{z}}(\bar{s})=\bar{\mathcal{C}}_{\rm ss}^{\tilde{\nu}_{*}\tilde{z};1}(\bar{s})+\bar{\mathcal{C}}_{\rm ss}^{\tilde{\nu}_{*}\tilde{z};2}(\bar{s}),
\end{equation}
with 
\begin{equation}\label{eq:1stconLT}
\bar{\mathcal{C}}_{\rm ss}^{\tilde{\nu}_{*}\tilde{z};1}(\bar{s})=\frac{\bar{s}C_{\rm ss}^{\tilde{\nu}_{*}\tilde{z}}(0) + [\xi 2C\,C_{\rm ss}^{\tilde{\nu}_{*}\tilde{\nu}}(0)_{\rm ss}+C_{\rm ss}^{\tilde{\nu}_{*}\tilde{z}}(0)]}{(\xi +\bar{s})(1+\bar{s})+\xi 2C}
\end{equation}
and
\begin{equation}\label{eq:2ndconLT}
\bar{\mathcal{C}}_{\rm ss}^{\tilde{\nu}_{*}\tilde{z};2}(\bar{s})=\frac{\xi 2C  X^4}{(1+2C)(\xi+1)} \frac{(\xi+\bar{s})(\xi+1)+2C\bar{s}}{[(\xi+\bar{s})(1+\bar{s})+\xi 2C]^2},
\end{equation}
both of which are terms of order $X^4$ in the small intracavity amplitude. Factorizing the denominator in Eqs. \eqref{eq:1stconLT} and \eqref{eq:2ndconLT} as 
\begin{equation}
(\xi +\bar{s})(1+\bar{s})+\xi 2C=(\bar{s}-\bar{\rho}_{-})(\bar{s}-\bar{\rho}_{+}),
\end{equation}
with $\bar{\rho}_{\pm}\equiv -\frac{1}{2}(\xi+1)\pm i \bar{G}$ and 
\begin{equation}\label{eq:Gdef}
\bar{G}\equiv \sqrt{\xi 2C-\frac{1}{4}(\xi-1)^2},
\end{equation}
we find the inverse Laplace transform of the two components as
\begin{equation}\label{eq:1stcondR}
\begin{aligned}
&C_{\rm ss}^{\tilde{\nu}_{*}\tilde{z};1}(\bar{\tau})=\exp\left[-\frac{(\xi+1)}{2}\bar{\tau}\right]  \Bigg\{C_{\rm ss}^{\tilde{\nu}_{*}\tilde{z}}(0)\cos(\bar{G}\bar{\tau})\\
&+\frac{\xi 2C\,C_{\rm ss}^{\tilde{\nu}_{*}\tilde{\nu}}(0)_{\rm ss}+[(1-\xi)/2]C_{\rm ss}^{\tilde{\nu}_{*}\tilde{z}}(0)}{\bar{G}}\sin(\bar{G}\bar{\tau})\Bigg\}
\end{aligned}
\end{equation}
and
\begin{equation}\label{eq:2ndcondR}
\begin{aligned}
&C_{\rm ss}^{\tilde{\nu}_{*}\tilde{z};2}(\bar{\tau})=\frac{\xi 2C X^4}{(1+2C)(\xi+1)} \frac{1}{2\bar{G}}\exp\left[-\frac{(\xi+1)}{2}\bar{\tau}\right] \\
&\times \Bigg\{\frac{(\xi+1)(\xi-1-2C)}{\bar{2G}} \left[\frac{\sin(\bar{G}\bar{\tau})}{\bar{G}}-\bar{\tau}\cos(\bar{G}\bar{\tau})\right]\\
& + (1+\xi+2C)\bar{\tau}\sin(\bar{G}\bar{\tau})\Bigg\}.
\end{aligned}
\end{equation}
The above two components alongside their sum are depicted in Fig. \ref{fig:components} for the region of collective strong coupling, in which spectral features with a clear offset from the empty-cavity resonance were observed in the experiment of \cite{Raizen1989}. We note that the frequency of the vacuum Rabi oscillation \cite{Carmichael1986} as defined in Eq. \eqref{eq:Gdef} in its scaled form, is also a function of the dissipation rates, with $G \equiv (\gamma/2) \bar{G}=g \sqrt{N}$ only for impedance-matching conditions ($2\kappa=\gamma$).

Let us here compare to the correlation function associated with the incoherent spectrum of the transmitted light [see Sec. 15.2.6 of \cite{CarmichaelQO2}], separated once more in two components, with Laplace transform
\begin{equation}\label{eq:LTtrlight}
\bar{\mathcal{C}}_{\rm ss}^{\tilde{z}_{*}\tilde{z}}(\bar{s})=\bar{\mathcal{C}}_{\rm ss}^{\tilde{z}_{*}\tilde{z};1}(\bar{s})+\bar{\mathcal{C}}_{\rm ss}^{\tilde{z}_{*}\tilde{z};2}(\bar{s}),
\end{equation}
where
\begin{equation}\label{eq:LTtrlight1}
\frac{\bar{\mathcal{C}}_{\rm ss}^{\tilde{z}_{*}\tilde{z};1}(\bar{s})}{X^4}= \frac{4C^2(2+\xi+2C)}{(1+2C)^2(\xi+1)^2} \frac{1+\xi+\bar{s}}{(\xi+\bar{s})(1+\bar{s})+\xi 2C}
\end{equation}
and 
\begin{equation}\label{eq:LTtrlight2}
\frac{\bar{\mathcal{C}}_{\rm ss}^{\tilde{z}_{*}\tilde{z};2}(\bar{s})}{X^4}=\frac{4C^2 \xi}{(1+2C)(\xi+1)}\frac{\xi(\xi-2C+\bar{s})}{[(\xi+\bar{s})(1+\bar{s})+\xi 2C]^2}.
\end{equation}
Our interest here is with the component given by Eq. \eqref{eq:LTtrlight2} \textemdash{also} proportional to $X^4$ as compared to the analogous term given by Eq. \eqref{eq:2ndconLT} \textemdash{responsible} for the squared-Lorentzian distribution in the spectrum of the transmitted light [obtained from $\bar{\mathcal{C}}_{\rm ss}^{\tilde{z}_{*}\tilde{z};2}(\bar{s})$ for $\bar{s}=-i2(\omega-\omega_0)/\gamma$]. The corresponding contribution to the correlation function for the cavity field is
\begin{equation}\label{eq:trlight2R}
\begin{aligned}
&C_{\rm ss}^{\tilde{z}_{*}\tilde{z};2}(\bar{\tau})=X^4\frac{4C^2 \xi}{(1+2C)(\xi+1)} \frac{1}{2\bar{G}}\exp\left[-\frac{(\xi+1)}{2}\bar{\tau}\right] \\
&\times \Bigg\{\frac{\xi(\xi-1-4C)}{\bar{2G}} \left[\frac{\sin(\bar{G}\bar{\tau})}{\bar{G}}-\bar{\tau}\cos(\bar{G}\bar{\tau})\right]\\
& + \xi \bar{\tau}\sin(\bar{G}\bar{\tau})\Bigg\}.
\end{aligned}
\end{equation}
The sum of two corresponding components of a light-matter correlation and the cavity-field autocorrelation, $[C_{\rm ss}^{\tilde{\nu}_{*}\tilde{z};2}(\bar{\tau}) + C_{\rm ss}^{\tilde{z}_{*}\tilde{z};2}(\bar{\tau})]/X^4$ obtained from Eqs. \eqref{eq:2ndcondR} and \eqref{eq:trlight2R}, is compared to the component $C_{\rm ss}^{\tilde{\nu}_{*}\tilde{z};2}(\bar{\tau})/X^4$ alone in Fig. \ref{fig:compRes} as we approach the many-atom strong-coupling limit of absorptive bistability [$\xi 2 C\gg (\xi+1)^2/4$]. Upon a further increase of the parameter $\xi 2C$, the two curves coincide.
\begin{figure}
\centering
\includegraphics[width=0.4\textwidth]{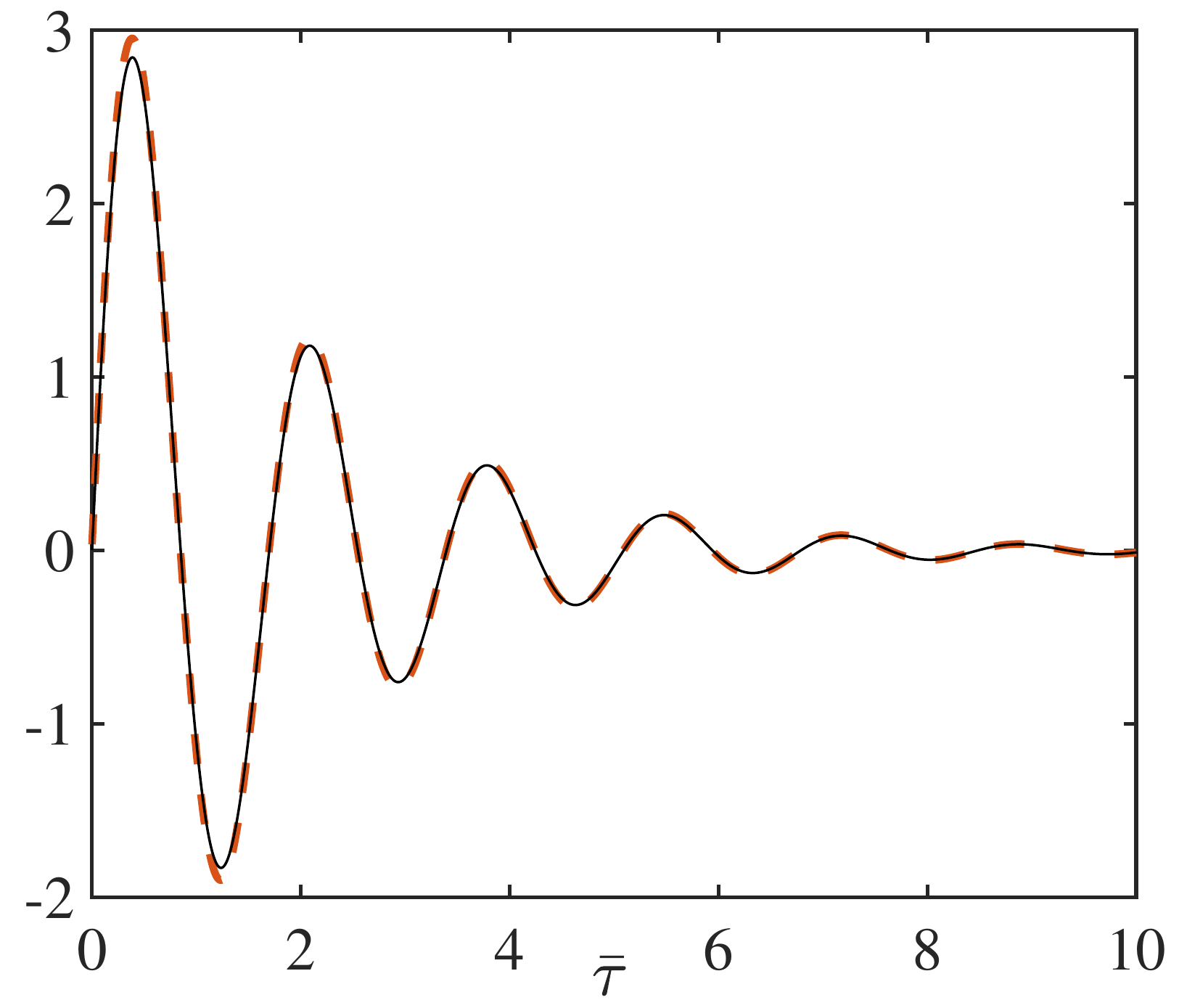}
\caption{{\it Correlations with a Lorentzian distribution in the good-cavity limit.} The component $C_{\rm ss}^{\tilde{\nu}_{*}\tilde{z};1}(\bar{\tau})/X^4$ from Eq. \eqref{eq:1stcondR} is plotted in a solid black line, superimposed on the correlation $C_{\rm ss}^{\tilde{\nu}_{*}\tilde{z}_{*}}(\bar{\tau})/(-X^2)$ obtained from Eq. \eqref{eq:corrX2} and depicted by a dashed orange line. We use the same parameters as in Fig. \ref{fig:components}, except for $\kappa/(2\pi)=0.18\,$ MHz, giving $C \approx 194$ and $\xi \approx 0.04$.}
\label{fig:compLorentzian}
\end{figure}

On the other hand, the correlation function obtained from Eq. \eqref{eq:LT_b}, 
\begin{equation}\label{eq:corrX2LT}
\bar{\mathcal{C}}_{\rm ss}^{\tilde{\nu}_{*}\tilde{z}_{*}}(\bar{s})=- X^2\frac{\xi 2C}{(\xi+1)(1+2C)} \frac{\bar{s}+\xi+2(C+1)}{(\xi +\bar{s})(1+\bar{s})+\xi 2C},
\end{equation}
as
\begin{equation}\label{eq:corrX2}
\begin{aligned}
C_{\rm ss}^{\tilde{\nu}_{*}\tilde{z}_{*}}(\bar{\tau})&=-X^2\frac{\xi 2C}{(\xi+1)(1+2C)} \exp\left[-\frac{(\xi+1)}{2}\bar{\tau}\right] \\
&\times\Bigg[\cos(\bar{G}\bar{\tau}) + \frac{4C+\xi+3}{2\bar{G}}\sin(\bar{G}\bar{\tau}) \Bigg],
\end{aligned}
\end{equation}
dominates in the weak-excitation limit, being of order $X^2$ instead of $X^4$. In the good-cavity limit ($\xi \ll 1$), we find that $C_{\rm ss}^{\tilde{\nu}_{*}\tilde{z}_{*}}(\bar{\tau})/X^2$ cancels $C_{\rm ss}^{\tilde{\nu}_{*}\tilde{z},1}(\bar{\tau})/X^4$, as depicted in Fig. \ref{fig:compLorentzian}. The same order is shared by the atom-atom correlation
\begin{equation}\label{eq:atatCorr}
\bar{\mathcal{C}}_{\rm ss}^{\tilde{\nu}_{*}\tilde{\nu}_{*}}(\bar{s})=-\frac{X^2}{(\xi+1)(1+2C)} \frac{(1+\xi+2C)\bar{s}+\xi(\xi+1)}{(\xi +\bar{s})(1+\bar{s})+\xi 2C},
\end{equation}
as well as by the field-field correlation $\bar{\mathcal{C}}_{\rm ss}^{\tilde{z}_{*}\tilde{z}_{*}}(\bar{s})$ which determines the second-order coherence properties of the transmitted light \cite{Carmichael1986}. 

The cross-correlation given by Eq. \eqref{eq:corrX2} is to be compared with another anomalous correlation given by Eq. (15.128) of \cite{CarmichaelQO2},
\begin{equation}\label{eq:corrQO2LT}
\bar{\mathcal{C}}_{\rm ss}^{\tilde{z}_{*}\tilde{\nu}_{*}}(\bar{s})=-X^2\frac{\xi 2C}{(\xi+1)(1+2C)} \frac{\xi+\bar{s}-2C}{(\xi+\bar{s})(1+\bar{s})+\xi 2C},
\end{equation}
inverse-transforming to
\begin{equation}\label{eq:corrQO2}
\begin{aligned}
C_{\rm ss}^{\tilde{z}_{*}\tilde{\nu}_{*}}(\bar{\tau})&\equiv N\lim_{\bar{t} \to \infty}\braket{\Delta \tilde{\bar{a}}^{\dagger}(\bar{t}) \Delta\tilde{\bar{J}}_{+}(\bar{t}+\bar{\tau})}\\
&=-X^2 \frac{\xi 2C}{(\xi+1)(1+2C)} \exp\left[-\frac{(\xi+1)}{2}\bar{\tau}\right] \\
&\times\Bigg[\cos(\bar{G}\bar{\tau}) + \frac{\xi-4C-1}{2\bar{G}}\sin(\bar{G}\bar{\tau}) \Bigg].
\end{aligned}
\end{equation}
The two functions share the same initial value, $C_{\rm ss}^{\tilde{z}_{*}\tilde{\nu}_{*}}(0)=C_{\rm ss}^{\tilde{\nu}_{*}\tilde{z}_{*}}(0)$, as expected from the symmetry of the steady-state covariance matrix $\boldsymbol{C}_{\infty}$, but differ in their dynamical evolution. In the many-atom strong-coupling limit asymptotically defined by  $\xi \to 0$, $C \to \infty$, with $\xi 2 C \gg 1$ remaining constant, the sum of the two correlators,
\begin{equation}
C_{\rm ss}^{\tilde{z}_{*}\tilde{\nu}_{*}}(\bar{\tau})+C_{\rm ss}^{\tilde{\nu}_{*}\tilde{z}_{*}}(\bar{\tau})\approx -2 X^2 \xi \exp\left(-\frac{\bar{\tau}}{2}\right)\cos(\sqrt{\xi 2C}\bar{\tau}),
\end{equation}
tends to zero as $\xi \to 0$. Their difference, however, evaluating to
\begin{equation}
\begin{aligned}
&C_{\rm ss}^{\tilde{z}_{*}\tilde{\nu}_{*}}(\bar{\tau})-C_{\rm ss}^{\tilde{\nu}_{*}\tilde{z}_{*}}(\bar{\tau})\\
&\approx 2 X^2 \sqrt{\xi 2C} \exp\left(-\frac{\bar{\tau}}{2}\right)\sin(\sqrt{\xi 2C}\bar{\tau}),
\end{aligned}
\end{equation}
does not vanish as long as atomic coherence is maintained; this is a sign of the competition between a restricted bandwidth in the communication channel across individual atoms, $2\kappa/\gamma \to 0$, and the collective strong coupling of the atomic ensemble to the intracavity field. A photon emitted by the atomic ensemble is likely to be captured and re-emitted multiple times before its ultimate escape through the cavity mirrors.

In this Letter, we have derived analytical expressions for the atom-field correlation functions in the weak-excitation limit of absorptive optical bistability in the small-noise approximation of Gaussian fluctuations. We have shown that certain correlations are of order $X^2$ in the intracavity excitation strength, and dominate over those corresponding to the squared Lorentzian terms in the spectral distribution of the transmitted light, which are of order $X^4$. Anomalous correlations in the cavity-field amplitudes of order $X^2$ are responsible for squeezing and, through their negative sign, produce photon antibunching for the forwards-scattered light [$g_{\rightarrow}^{(2)}(0)-1<0$] alongside a negative spectrum of squeezing as first-order corrections in $N^{-1}$ (see Sec. 15.2.7. of \cite{CarmichaelQO2} and \cite{Carmichael1986} for more details). Despite their different orders of magnitude, however, the contributions scaled by the appropriate powers of $X$ are related to each other. Namely, the scaled components of different atom-field correlations with a Lorentzian spectral distribution tend to opposite values in the many-atom strong-coupling regime. At the same time, for increasing values of $\xi 2C$, the sum of those components contributing at order $X^4$ oscillates in coincidence with the transmitted-light correlator having a squared Lorentzian distribution. 

The simplest way to access collective atomic emission is to couple the atomic ensemble to a second cavity in addition to the primary resonator realizing the bistable absorber, as suggested by \cite{CarmichaelPC}. Let the coupling to the second cavity be weak and the linewidth of that cavity be large enough to justify an adiabatic elimination of the supported resonant mode. This configuration provides a collective decay channel \cite{Carmichael1983} via the second cavity mode. Since the photon flux escaping via that channel is sufficiently low, we neglect its effect on the overall system dynamics. Monitoring the second channel via homodyne detection reveals a negative source-field spectrum of squeezing $S(\omega)\propto {\rm Re}[\bar{\mathcal{C}}_{\rm ss}^{\tilde{\nu}_{*}\tilde{\nu}_{*}}(-2i \omega/\gamma)]$ when the local oscillator is in phase with the mean collective atomic polarization, while combining the fields transmitted from the two cavities yields the cross-correlations of light-matter interaction in absorptive optical bistability.

\begin{acknowledgments}
I am grateful to Prof. H. J. Carmichael for instructive discussions and guidance. I also wish to acknowledge financial support by the Swedish Research Council (VR) alongside the Knut and Alice Wallenberg foundation (KAW).
\end{acknowledgments}

\begin{center}
 *****
\end{center}

\end{document}